# Phase manipulation of terahertz waves by work function engineering in metal-graphite structures

Muhammad Irfan,[1] Soo Kyung Lee,[1] Jong-Hyuk Yim,[1] Yong Tak Lee,[1,2] and Young-Dahl Jho[1, a)]

[1)] *School of Information and Communications, Gwangju Institute of Science and Technology, Gwangju 500-712, South Korea*

[2)] *Advanced Photonics Research Institute, Gwangju Institute of Science and Technology, Gwangju 500-712, South Korea*

(Dated: 17 February 2016)

We manipulate the transient terahertz (THz) waves emitted from metal-graphite interfaces, where potential barriers were formed because of work function differences. To adaptively control the phase of the THz waves, two distinct groups of metals were evaporated on *n*-type doped highly oriented pyrolytic graphite (HOPG): group A, which consisted of Pt, Au, and Ag with work functions larger than that of HOPG and group B, which consisted of Al and Ti with work functions smaller than that of HOPG. The phase of the transient THz lineshapes from group A was opposite to that of group B under infrared laser excitation, which is indicative of opposite band bending and concomitant interfacial doping for ambipolar transport at the metal-graphite junctions. The amplitude of the THz waves could be further substantiated by the work function differences and modified minority carrier mobilities at the depletion regions.

In the field of terahertz (THz) science and technology, engineering the functionalities of THz waves such as the amplitude, phase, spectrum and directionality is of great importance for various potential applications such as THz imaging[1], coherent control of molecular dynamics[2] and remote sensing.[3] The emitted THz waves from different materials, which yields information not only on the amplitude but also on the phase, provides insight into the ultrafast dynamics of non-equilibrium carrier transport with a femtosecond time scale e.g., direction of band bending and surge currents as compared to conventional methods.[4,5] A rich variety of approaches have been demonstrated by changing material parameters and ambient conditions. Early studies indicated that polarity reversal of THz radiation could be achieved as a function of pump laser wavelength around the band edge in bulk

---

a) Electronic mail: jho@gist.ac.kr

semiconductors[6] or as a function of temperature in dilute magnetic semiconductors.[7] The polarity reversal of THz single-cycle pulses has been observed as result of the Gouy phase shift passing through a focal plane.[8] The opposite doping type was another source for THz polarity flip in materials such as GaAs, InP,[9] and graphite[10] because of the reversal of the direction of the surface depletion field. The polarity of the emitted THz waves from air plasma was completely controlled by the relative phase between the linear and nonlinear transitions.[11] Recently, polarity reversal of THz waveforms has been observed by switching the helicity of circular polarization pulses in NiO crystals[12], changing the magnetic field direction or bias voltage in graphite/photoconductive antenna[13–15], ultrafast laser heating of edges in thermoelectric materials[16], or changing the film thickness of InAs.[17] Moreover, the temporal evolution of transient THz lineshapes can be precisely manipulated using an optical pulse shaper.[18]

From a materials perspective, carbon-based materials have recently generated great interest in the THz community. Since the advent of graphene as one of the building blocks in functional THz materials[19], THz radiations from graphene[20] and graphite have also been reported both in the far-[10,13] and near-field regime.[21] However, prospective heterostructures such as metal-graphite interfaces have not been explored in terms of THz generation via light-matter interactions, even though THz features at the semiconductor-metal interfaces have long been reported.[22] In particular, the selection of metals to control the magnitude and direction of the depletion field at the metal-graphite interface could provide further insight into the feasibility of work function engineering in related structures. In this work, we demonstrate how the modified work function at the potential barriers between metal and graphite can affect the phase of THz waves. To gain a better understanding of the underlying mechanisms involved, we further substantiate the effect of the azimuthal angle, excitation power, and detection geometries.

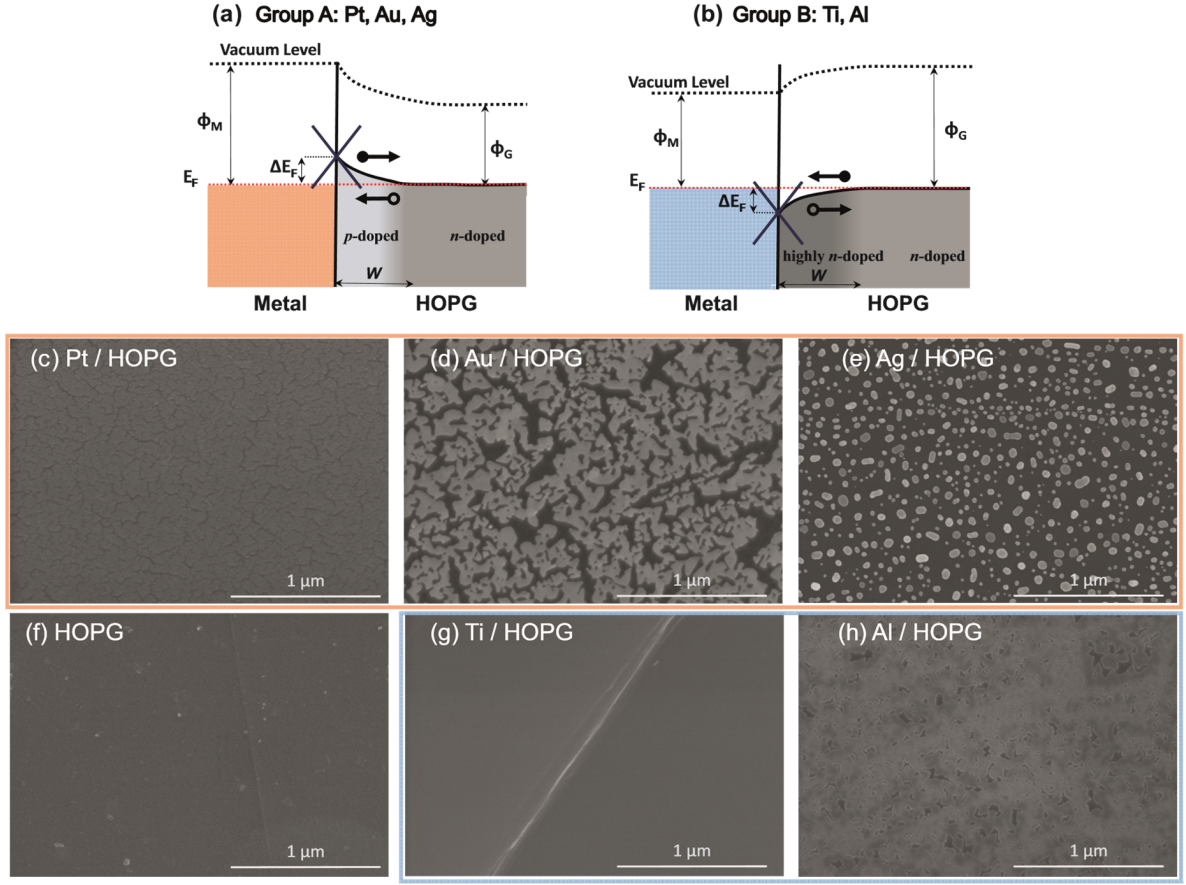

FIG. 1. Schematic illustration of the band structure for (a) group A and (b) group B, exhibiting *p*-type doping and highly *n*-type doping near interfaces. The parameters $\varphi_M$ ($\varphi_G$) and $\Delta E_F$ correspond to the metal (HOPG) work function and Fermi-level shift, respectively, and $W$ denotes the depletion region. Electrons and holes in energy band dispersions at the K-valley are represented with filled and open circles with arrows, respectively indicating the drift motion. SEM images of (c-e) group A, (g,h) group B (outlined in golden and blue, respectively), and (f) HOPG.

Highly oriented pyrolytic graphite (HOPG) was prepared by repeatedly peeling off the surface layers. Metal layers (Pt, Au, Ag, Ti, Al) of 5 nm thickness were deposited on freshly cleaved 1 mm thick HOPG using electron-beam evaporation (for Pt, Au, Ag, Ti) or thermal evaporation (for Al), followed by annealing for 10 min at 200°C in a nitrogen environment.

The samples were divided into two groups based on the metallic work function ($\varphi_M$):[23] Pt (5.3 eV), Au (4.8 eV) and Ag (4.64 eV) on HOPG (4.6 eV), referred to as group A with $\varphi_M$ values greater than that of HOPG, and Ti (4.33 eV) and Al (4.25 eV) on HOPG as group B with $\varphi_M$ values less than that of HOPG. Figure 1(a, b) displays the band diagrams at the interfaces of the two groups. The doping of graphene or graphite by metals can be explained by charge transfer between the metal and graphite due to difference in work functions.[24,25] The dipole formed due to charge redistribution at the interface results in Fermi-level shift relative to Dirac-point. A downwards (upwards) shift in Fermi-level for group A (B) indicates that holes (electrons) are donated by the metal to graphite which becomes p-type (n-type) doped. The distinct band bending between groups highlights the carrier drift near the interfaces as a major THz emission mechanism among various transient transport processes. In contrast, the transient diffusive contribution to THz radiation can be ignored because of the similar mobility of electrons and holes[26] despite the thin skin-depth of ~20 nm under 800 nm excitation (compared with the diffusion length of ~500 nm)[10]. The drift currents resulting from photo-carriers can be modified by the work function difference at the interface $\varphi_M - \varphi_G \equiv \Delta\varphi$, possibly leading to a distinct phase of emitted THz waves.

The metals Pt, Au, Ag and Al bond weakly to graphite and result in a physisorptive interface, whereas Ti bonds strongly and results in a chemisorptive interface.[27] Figure 2 presents scanning electron microscopy (SEM) images of islands of Au and Ag, whereas Pt, Ti and Al are absorbed better into the surface and form smoother films. The morphology of metals strongly depends on their adsorption energy with respect to the substrate. The adsorption of transition metals on graphite has been studied extensively, and the adsorption energies for Pt, Au, Ag, Ti and Al with respect to graphite are 1.86, 0.49, 0.28, 3.3, and 1.6 eV, respectively.[28] Of these metals, Ti has the largest adsorption energy to graphite, resulting in the smoothest film development. In contrast, Ag has the lowest adsorption energy, inducing aggregation features and nanoparticle formation. The coverage area for Pt/HOPG (80%), Au/HOPG (53%), Ag/HOPG (23%), Ti/HOPG (99%), and Al/HOPG (82%) were estimated by gray thresholding.

The photo-carriers were abruptly generated under a Ti-sapphire laser excitation centered at 800 nm, with a pulse width 150 fs. Conventional THz time-domain spectroscopy (THzTDS) was employed at 300 K with two different detection geometries to characterize whether the phase of THz waves could be associated with the radiation out-coupling efficiencies.

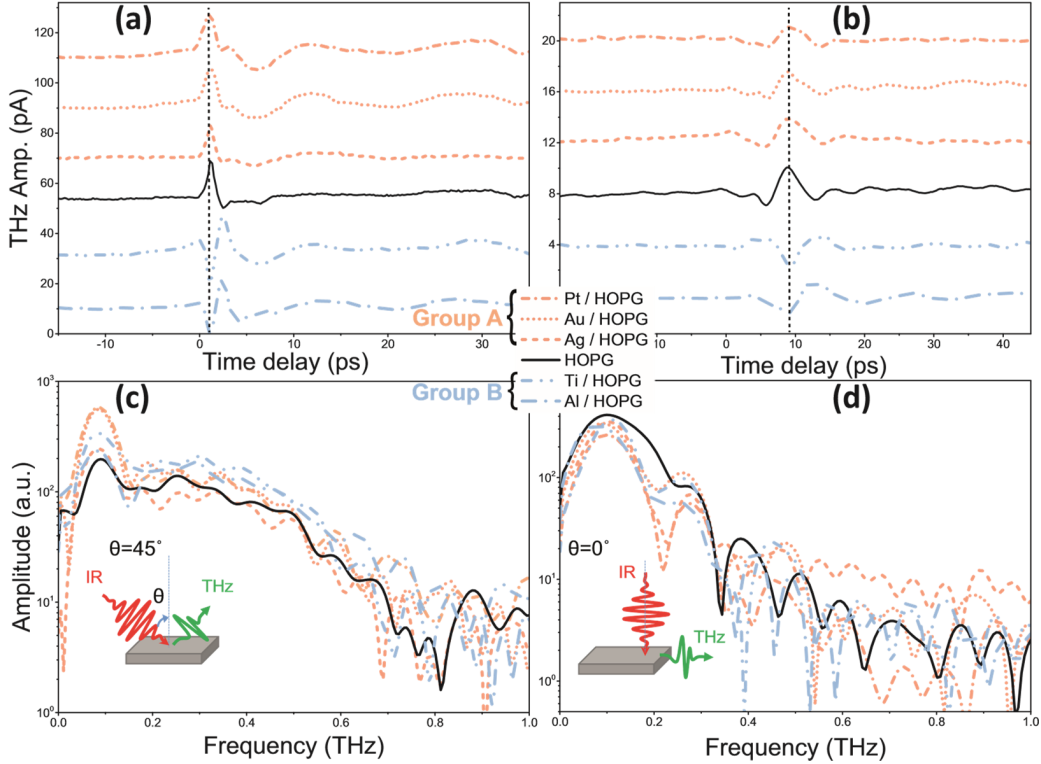

FIG. 2. Time-domain signals of THz waves and corresponding amplitude spectra from metal-HOPG structures in (a, c) conventional reflection geometry and (b, d) edge detection geometry. The insets show the experimental geometries of THz wave radiation (green arrow) under IR excitation (red arrow). The spectral peaks and troughs in the edge detection geometry (b,d) results from destructive interference across the illumination spot.

The infrared (IR) laser source was focused onto a spot the size of 300 $\mu$m and the incident angle $\theta$ was 45° in the reflection geometry, whereas $\theta$ was 0° (normal incidence) in the edge detection geometry. The pump fluence at the metal-graphite interfaces was estimated by subtracting the reflected fluence from the incident value and was held below 0.2 $\mu$J/cm$^2$ to avoid both the screening of the depletion field by the photo-carriers and nonlinear effects including optical rectification.[29] Following femtosecond pulse pumping, the generated THz wave from the samples was guided using a pair of off-axis parabolic mirrors and focused on photo-conductive antenna whose sensitivity was optimized at 1 THz.

These work function-dependent drift features were experimentally demonstrated by tracing the lineshapes of THz waves as shown in Fig. 2. Both in the reflection geometry [Figs. 2(a,c)] and edge detection geometry [Figs. 2(b,d)], the emitted THz fields from group A exhibited the same polarity as those from HOPG ; however, the THz pulses emitted from group B revealed a polarity flip. The abrupt photo-carrier generation is expected to generate polarized electron-hole pairs during the optical excitation in the depletion region, followed by the translational carrier transport, governed by the drift near the interfaces in our structures at early time delays. The opposite polarity of the THz waves in the time domain, therefore, suggests that the acceleration direction of polarized Hertzian dipoles[22] is reversed between different groups near the depletion region. We also note that the band bending at the interface for group A is expected to lie along the same direction as that of the air/nHOPG interface, as previously revealed based on the Fermi-level pinning effect at the surface charge layer (SCL).[10] In contrast, Hertzian dipole acceleration along the reversed direction in group B was achieved near the interface because of the oppositely tilted potential gradients as illustrated in Figs. 1(a, b).

To further clarify the contribution from the carrier transport along the *c*-axis to the phase changes, Fig. 2(b) presents the time-domain THz radiation patterns in the edge detection geometry. In such geometry, the THz emission from the carrier transport along the basal plane and nonlinear effects could be avoided. Consequently, the THz pulses emitted from transient dipoles along *c*-axis, presumably influenced by inhomogeneous surface morphology during metal evaporations, should form distinct directional radiation patterns as precisely investigated in low band gap semiconductors.[30,31] The line-of-sight propagation of THz waves therein has been modeled, incorporating anisotropic radiation power proportional to $\sin^2 \theta$ and perpendicular to dipole axis. Remarkably, the phase of the THz waves from group A is opposite to that of group B, which confirms the effect that the $\Delta\varphi$ has on the phase of THz lineshapes. However, the THz intensity in this geometry was relatively low and temporally broadened, depending on its position from the edge, as discussed elsewhere.[32,33] Considering the smaller laser spot size (~ 300 $\mu$m) and relatively large excitation position away from the edge (~ 1 mm) compared with the THz wavelength, the THz waves emitted from all places added together coherently. These laterally propagation THz waves were subjected to the strong attenuation and temporal broadening which can be evaluated based on the refractive index, absorption coefficient and phase velocity. The narrow spectrum in edge detection geometry [Fig. 2(d)] can be associated with the dispersion behavior of materials in the THz regime.[34]

In addition to the transient photocurrent, the effect of nonlinear optical processes could be experimentally examined. In this regard, the optical rectification is considered as another possible mechanism for the generation of broadband THz pulses in materials with significant second-order susceptibility $\chi^{(2)}$ despite the graphite structure having a center of inversion symmetry that prohibits $\chi^{(2)}$ process from the bulk.[35] However, whether the inversion symmetry is broken by the depletion fields, intercalation, or stacking faults can be determined based on the nonlinearity-induced variations of THz amplitudes with azimuthal angle Ψ in the reflective geometry, as illustrated in Fig. S1(a).[51] Conventionally, the $\chi^{(2)}$ process leads to a well-defined periodicity in axial measurements such as four-fold symmetry in (100) InAs and six-fold symmetry in (111) InAs.[29] However, THz amplitude variations in Fig. S1 (a) exhibit negligible Ψ dependence. This is due to fact that the photo-generated transient current is rotationally symmetric along the surface normal (*c*-axis) and the translation symmetry on the graphite side remains preserved. Figure S1 (b) shows the influence of the excitation fluence on THz emission. For all the samples, the detected THz amplitude mostly increased linearly with the pump fluence. There is no evidence of saturation, which suggests that the photo-carrier screening is too small to compensate for the depletion fields. We note that our fluence (~μJ/cm²) did not reach the quadruple second order nonlinear regime reported in graphite (~mJ/cm²).[36]

To explain our experimental observations, the THz field amplitude due to diffusive and drift motion in the depletion field is simplified by adjusting relevant material parameters:[39]

$$E_{drift} = \frac{S\mu e^2 \triangle n N (e^{-\alpha W} + \alpha W - 1)}{4\pi\epsilon_0^2 \epsilon_r \alpha^2 c^2 \tau r}, \quad (1)$$

$$E_{diffusion} = \frac{Sk_B T \mu}{4\pi\epsilon_0 c^2 \tau r (1+b)} \left\{ \frac{bN}{1+b} \times \ln[1 + \frac{\triangle n(1+b)}{bN}] - \triangle n \right\} \quad (2)$$

$$E_{far} = E_{drift} \pm E_{diffusion} \quad (3)$$

Where $\epsilon_0(\epsilon_r)$ is the permittivity in free space (graphite~ 6.6)[40], $\alpha$ is the absorption coefficient (~ $5 \times 10^5$ cm$^{-1}$), $e$ is the elementary charge of the electron, $4n$ is the photo-carrier density at the surface, $\mu$ is the mobilities of electrons or holes, $\tau$ is the duration of pulse, $r$ is the distance between emission source and detector, $N$ is the carrier concentrations, $W$ is the depletion width, $b$ is the ratio of mobilities of electrons and holes, $k_B$ is the Boltzmann constant, and $T_e$ the carrier temperature obtained from residual photon energy. The carrier concentration and mobility were measured using Hall measurements in the Van der Pauw configuration. The corresponding depletion width, carrier concentration and photo-carrier density for all the samples are listed in Table S1.[51]

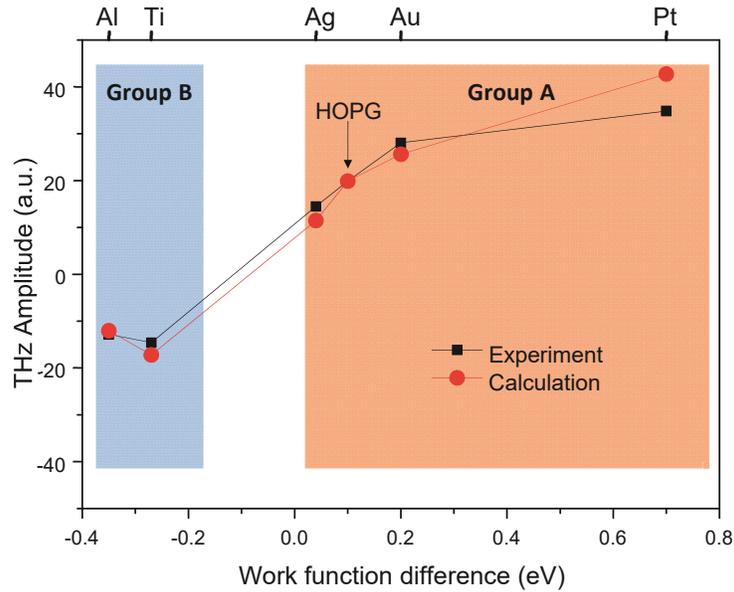

FIG. 3. THz amplitude in reflective geometry and calculated radiated THz field plotted as a function of the work function difference of metal-HOPG structures.

The depletion width ($W$) at the metal-graphite interface is expressed as follows:[41] $W = 2\sqrt{\frac{\epsilon_0 \epsilon}{eN}} \times ln[\frac{\varphi_s e}{k_B T}]$, where $N$ is the carrier density at the Fermi level, $T$ is temperature, and $\phi_s$ is the potential drop in SCL (the work function disparity between metal and graphite $\Delta\varphi$) computed from band bending as illustrated in Fig. 1. The THz far-field due to carrier diffusion ($E_{diffusion}$) is negligibly small as compared with carrier drift contribution ($E_{drift}$) based on the parameters in Table S1. Accordingly, the photo-generated carriers are accelerated in the depletion layer at the interface and concomitantly emit THz pulses where electric field polarity is coherently directed depending on the direction of band bending.

To gain further insight into the key factors associated with the THz field amplitude, we compared the THz amplitudes (black scattered line) to the calculated values ($E_{far}$; blue scattered line) based on Eq. (3) for each sample as shown in Fig. 3. In general, we note that the THz amplitude tends to increase with $\Delta\varphi$ whereas slight reduction was observed in Pt. The origins of this decline of the THz amplitude at relatively large $\Delta\varphi$ could be associated with the optically induced space charge, thus, compensating for the built-in field and saturating the electronic transports.[42] The waning amplitude in Al originated from the relatively lower mobility and photo-carrier density as confirmed by the suppressed calculated value in Table S1.**[51]**

The calculated values for the THz field are consistent with the experimental data considering ambipolar carrier transport at the interface where the behavior of THz radiation processes are dominated by the photo-generated minority carriers.[43] The magnitude of emitted THz radiation due to carrier drift [Eq. 1] is linearly dependent on the mobility of carriers in the depletion region which is calculated from the mobility of the respective minority carriers. Since carrier mobility is inversely proportional to the effective mass and the effective mass of holes in graphite is known to be almost two time higher than that of electrons, the mobility of electrons is approximately two times that of holes.[44,45] This finding clearly explains the decreasing THz amplitude in group B where behavior of excess carriers is determined by parameters of minority carriers holes. Furthermore, the origin of the discrepancies between the experimental and calculated values can possibly be attributed to the interfacial defects[35], additional chemical compositions and variation of extraction efficiencies[46] among metal/HOPG samples with different effective refractive indices, which will further discussed elsewhere. In this way, $\Delta\varphi$ was not sufficient to solely determine the THz magnitude, although the sign of $\Delta\varphi$ was directly associated with the phase of THz fields and the simplified estimation was qualitatively consistent with our experimental findings.

Additional noticeable features of THz waves in structures could be discussed in terms of plasmons induced by the rough metal surface under the femtosecond laser excitation and concomitant THz radiation from plasmonic nanoparticle arrays and thin metal films.[37,38] The random morphology of the agglomerated islands (Au, Ag) or fractal like structures of films (Pt, Al) in Fig. 1 suggests the potential existence of hot spots for localized surface plasmons and results in enhanced THz amplitude.[48] The excitation of these surface plasmons could also contribute to the enhancement of THz radiation through depletion field induced second order non-resonant optical rectification.[49] Moreover, internal photoelectric emission of charge carriers from metal into the graphite results in additional transient currents.[50] The photon drag effect in graphene layers near interface can also lead to enhancement of the THz emission.[20] Sometimes, the pump light can be confined to a very thin layer at the metal-graphite interface by Fabry-Perot cavity resonance, leading to increased optical absorption in the depletion region.[49] The coverage area could be increased by using thicker metal films for cavity-enhanced absorption, together with optimized HOPG thickness for the Fabry-Perot interference, to comprehensively intensify the THz generation.

We experimentally characterized THz electromagnetic waves emitted from metal-graphite junctions. We interpreted the opposite phase between different sample groups as a demonstration of work function engineering in controlling the phase of THz radiation. We also demonstrated that azimuthal angle change do not affect the THz phase or the amplitude with the symmetries being preserved in the *c*-plane in our ambient conditions. Such manipulation of THz pulses could find practical application in functional THz devices such as modulators for communication systems and active elements for THz sources.

This work was supported by the National Research Foundation of Korea (NRF) funded by the Ministry of Education, Science, and Technology (NRF-2013-068982).

# Phase manipulation of terahertz waves by work function engineering in metal-graphite structures

Muhammad Irfan,[1] Soo Kyung Lee,[1] Jong-Hyuk Yim,[1] Yong Tak Lee,[1,2] and Young-Dahl Jho[1, a]

[1] School of Information and Communications, Gwangju Institute of Science and Technology, Gwangju 500-712, South Korea

[2] Advanced Photonics Research Institute, Gwangju Institute of Science and Technology, Gwangju 500-712, South Korea

(Dated: 17 February 2016)

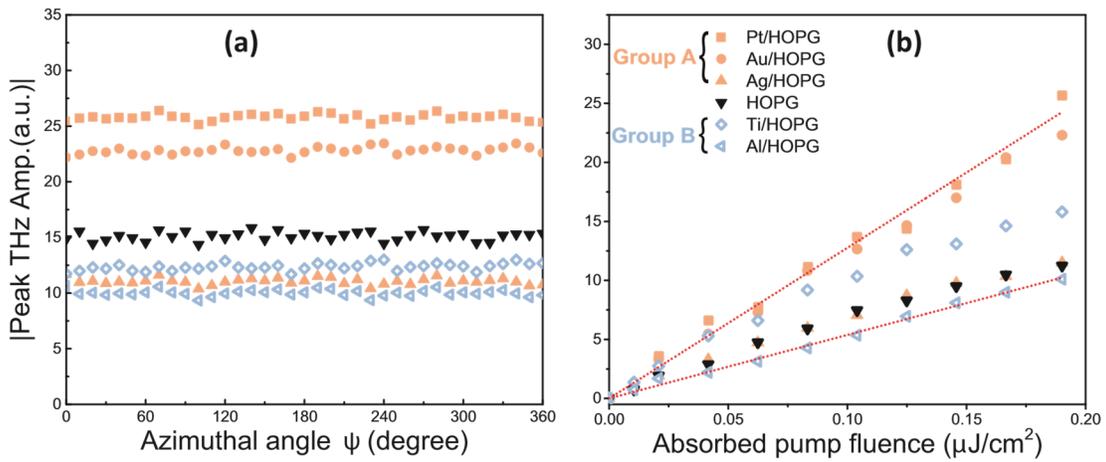

FIG. S1. (a) Azimuthal angle dependence of the THz field amplitude. (b) THz amplitude (scattered lines) measured as a function of excitation fluence, compared with the dotted linear lines.

a) Electronic mail: jho@gist.ac.kr

TABLE S1. Depletion width ($W$), mobility ($\mu$), carrier concentration ($N$), photo-carrier density ($4n$), and barrier height ($4\varphi$) for metal-graphite structures.

|  | Samples | Depletion width $W$ (nm) | Mobility $\mu$ (cm$^2$/V.s) | Carrier concentration $N$ ($10^{19}$ cm$^{-3}$) | Photo-carrier density $4n$ ($10^{14}$ cm$^{-3}$) | Barrier height $4\varphi$ (eV)[a] |
|---|---|---|---|---|---|---|
| Group A | Pt/HOPG | 27.2 | 820 | 1.62 | 1.879 | 0.7 |
|  | Au/HOPG | 14.88 | 920 | 2.29 | 1.891 | 0.2 |
|  | Ag/HOPG | 6.23 | 982 | 1.34 | 2.085 | 0.04 |
|  | HOPG | 9 | 1005 | 1 | 3.133 | 0.1 |
| Group B | Ti/HOPG | 20.02 | 1090 | 1.5 | 2.156 | -0.27 |
|  | Al/HOPG | 19.8 | 774 | 1.8 | 1.64 | -0.35 |

[a] Barrier height is estimated from metallic work function in previous work.[1]